# Breaking the general election effect: The impact of the 2020 US presidential election on China's economy and counter strategies

Junjie ZHAO[1]

[1] The Hong Kong University of Science and Technology (Guangzhou), Guangzhou City, Guangdong Province, China

**Abstract.** The study of US-China relations has always been a crucial topic in our economic development [4][5][7], and the US presidential election plays an integral role in shaping these relations. The presidential election is held every four years, and it is crucial to assess the impact of the 2020 election on China to prepare for the potential effects of the 2024 US presidential election on the Chinese economy [8][16][20]. To achieve this, we have gathered statistical data from nearly 70 years and analyzed data related to the US economy. We have classified the collected data and utilized the analytic hierarchy process [1][2][3] to evaluate the President's policy implementation. We have identified 15 evaluation indicators across five dimensions to assess the value of U.S. fiscal policy, monetary policy, employment and livelihood, regulatory policy, and the epidemic. These indicators form the basis of a comprehensive evaluation model that aims to determine the level of the U.S. economy. Due to the complex interplay between different factors and the numerous indicators involved, we utilized principal component analysis and factor analysis through the SPSS software to assess the degree of influence of each indicator on the economic level of the United States. This approach allowed us to obtain a comprehensive ranking of the indicators [6][9][11][33]. We then quantified the index data and employed the entropy weight method to calculate the weight of each index data. Finally, we used the weighted total score calculation to evaluate the economic status of the United States in a hierarchical manner after the election of Presidents Trump and Biden [15][18]. We optimized the index system by incorporating additional dimension indexes such as "foreign policy". We then crawled China's specific development data from 1990-2020 and substituted it into the model for analysis and evaluation. This enabled us to obtain detailed quantitative index data of the degree of influence [10][12][14]. Using estimates of the impact of different presidential candidates on China's economy, we established a comprehensive forecasting and evaluation model of the economic level based on the bi-level programming model. To address China's shortcomings in science and technology innovation, we recommend strengthening economic cooperation with developed countries, diversifying market development, and actively expanding the domestic market through feasible solutions [13][16][23][36].

**Keywords:** Principal Component Analysis, Factor Analysis, Entropy Method, Load Matrix, Comprehensive Evaluation Score Model, Bilayer Programming.

# 1 Assumptions

There are many factors that affect security check. To simplify the problems and make it convenient for us to simulate real- life conditions, we make the following basic assumptions, each of which is properly justified.

(1) Assume that we're doing it right and we're not doing it wrong when we go through a lot of data.
(2) Assume that the data collected are true and reliable.
(3) Assume that the indicators determined by us are independent of each other and do not overlap each other.
(4) Assume that the importance of indicators can be represented by the weight of indicators.

# 2 Symbol Descriptione

In the section, we use some symbols for constructing the model as follows:

| Symbol | Description |
|---|---|
| $e$ | Unit eigenvector |
| $m$ | Number of evaluated objects |
| $n$ | Number of evaluation factors |
| $minX_j$ | The minimum value of the JTH index |
| $maxX_j$ | The maximum value of the JTH index |
| $R_{ij}$ | Standardized index values |
| $X_{ij}$ | The index value of the ITH evaluation object under the JTH index |
| $w_j$ | The weight of the JTH index |
| $f_{ij}$ | Under the JTH index, the index value of ITH evaluation objects accounts for the total value of the JTH index of all evaluation objects |
| $t$ | Normalized orthogonal eigenvector matrices |
| $Zx$ | Matrix after variable standardization |
| $F$ | Composite factor score |

# 3 Data acquisition and analysis

According to the topic given in finance and trade, economic and financial governance, and other key development areas (such as COVID - 19 prevention measures, infrastructure, tax, environmental protection, medical insurance, employment, trade, immigration, education and other factors), by using the data from 1950-2020 and 70 statistical yearbook, and analysising to find out the data related to the U.S. economy.Through consulting baidu Encyclopedia, we know that the indicators to measure a country's economy include: GDP, CPI, PPI, unemployment rate and other indicators to measure comprehensively[1][2][3][7]. Therefore, the above data are used to represent the state of the national economy.

Web crawler is a program that can automatically grab web pages and extract web content. It is the information acquisition channel of search engines.It can simulate a browser to access network resources and automatically extract the required information.

Therefore, we adopt Python tools to collect and analyze data in the form of web crawler in Python3.7. We used Python3.7 to crawl the target web site (http://data.stats.gov.cn), baidu library, kuaiyi data and other web sites. Through the web sites, we found the relevant data of factor data required by the comprehensive assessment model of American economic level[4][5][6][8].

Some of the charts are shown below:

Gross domestic product (GDP) refers to the final outcome of the production activities of all resident units in a country (or region) calculated according to the national market price within a certain period. It is often recognized as the best indicator to measure the economic status of a country.Gross domestic product (GDP) is an important comprehensive statistical index in the accounting system, as well as the core index in China's new national economic accounting system, which reflects the economic strength and market size of a country (or region).

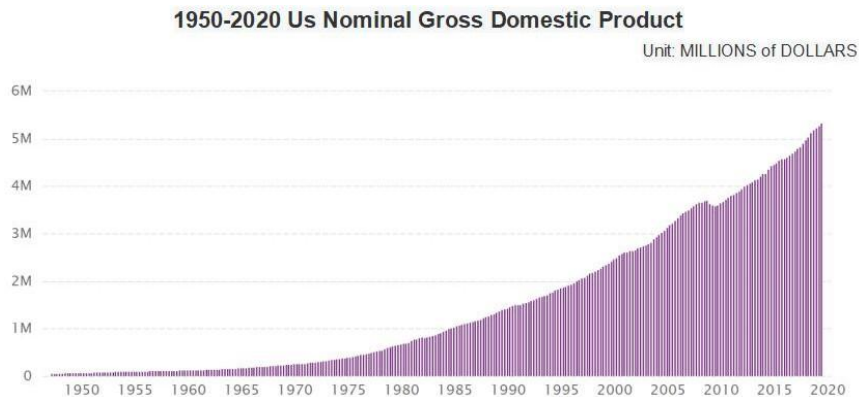

**Fig. 1.** 1950-2020 Us Nominal Gross Domestic Product

The Consumer Price Index is also known as the Consumer Price Index, or CPI.CPI is a macroeconomic indicator that reflects changes in the price level of consumer goods and services generally purchased by households.It is the relative number that measures the price level change of a group of representative consumer goods and services items over time in a specific period. It is used to reflect the change of the price level of consumer goods and services purchased by households. It is the change coefficient of the retail price of goods and services in a month[10][11][12].

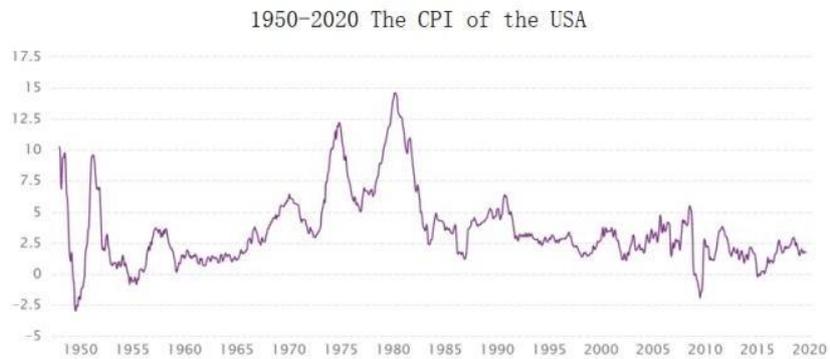

**Fig. 2.** 1950-2020 The CPI of the USA

Tax is a normative form in which the state, in order to provide public goods to the society, meet the common needs of the society and, in accordance with the provisions of the law, participate in the distribution of social goods, force and obtain fiscal revenue free of charge. Taxation is a very important policy tool[9][13][14].

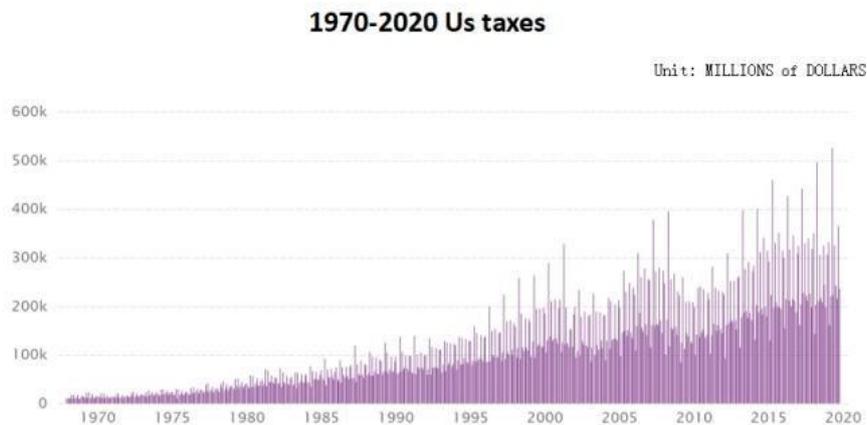

**Fig. 3.** 1970-2020 Us taxes

Unemployment rate refers to the number of unemployed workers who meet all employment conditions in a certain period of time. It is designed to measure the idle labor capacity and is the main indicator to reflect the unemployment status of a country or region[15][18].

Monthly changes in unemployment data may appropriately reflect economic development. There is a inverse relationship between unemployment rate and economic growth rate.In 2013, China for the first time released data on the surveyed unemployment rate[16][17][20].

## 4 The establishment of comprehensive evaluation model of American economic level

We categorize the data we collect and use the analytic hierarchy process to begin with the value of the President's policies. According to presidential policy decisions and literature review over the years, a total of 11 evaluation indicators can be extracted from the values of the five dimensions of U.S. fiscal policy[18][19][21], monetary policy, employment and livelihood, regulatory policy, and the epidemic, and an indicator system can be established.

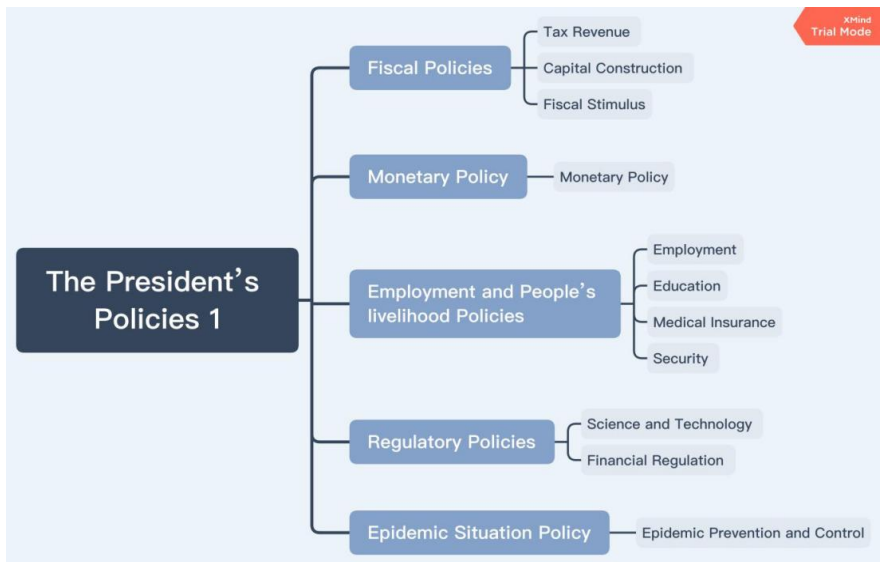

**Fig. 4.** American economy level evaluates index to classify

Since the dimensionality and the size of the index values among the collected factors will make the health assessment of the system difficult, it is necessary to standardize the relevant data in each influencing factor so as to facilitate data processing and calculation. "minimum-maximum standardization" (max − min) is adopted to process the data, which is to standardize the data and domesticate all the index values. The formula is as follows:

$$R_{ij} = (X_{ij} - minX_j)/(maxX_j - minX_j) \quad (1)$$

Among them, $minX_j$、 $maxX_j$ are the minimum and maximum values of the JTH index, and $R_{ij}$ are the standardized index values. According to the data, there are altogether 11 factors, but some of them are correlated, and the degree of influence of each factor on residents' health is also different.The principal component analysis (PCA) method[3] is selected first.Organizing the 11 factor data and listing them.Open the SPSS software, according to the "analysis - dimension reduction - factor analysis" the order of the open dialog box "factor analysis", the 11 variables move dialog box, the "description" child dialogs, "extraction" dialog box, "score" dialog, set the "options" dialog box, and then submit system operation.

According to the intermediate data, the first four major components have explained nearly 90.6% of the total variance, so the first four major components can be selected for analysis. Figure 6 is the lithotriptic diagram of the main components. Combined with the inflection point and characteristic root value of the characteristic root curve, it can be seen from the figure that the broken line slope of the first 4 main components is relatively steep, while the back surface gradually tends to be gentle. Therefore, it is appropriate to take the first 4 principal components.

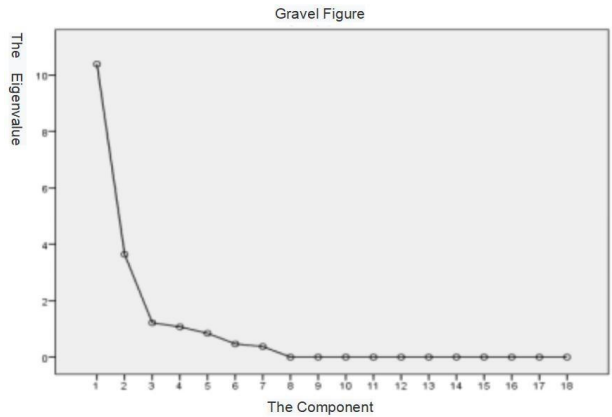

**Fig. 5.** Lithotripsy diagram of each component

Input the data from the factor loading matrix before rotation into the SPSS data editing window, and name the four variable names as $a_1$, $a_2$, $a_3$ and $a_4$ respectively, using the formula:

$$e_{ij} = a_{ij}/\sqrt{\lambda_i} \qquad (2)$$

Computing the normalized eigenvector, and computing $t_1$, $t_2$, $t_3$ and $t_4$. Since the correlation coefficient matrix is the starting point of the analysis, the variables of the original data should be standardized. After standardization, principal components are calculated in Excel, its expression is

$$y = Zx * t \qquad (3)$$

In the above equation, Zx is the normalized matrix of variables, and t is the normalized orthogonal eigenvector matrix. According to the variance percentage of principal component analysis, the comprehensive score is calculated as shown in Table 2. The first principal component accounts for 57.691%, the second principal component accounts for 20.225%, the third principal component accounts for 6.755%, and the fourth principal component accounts for 5.962%. The comprehensive score function is calculated, using the formula:

$$y = 0.57691y_1 + 0.20225y_2 + 0.06755y_3 + 0.05962y_4 \quad (4)$$

Then use factor analysis method for analysis.The operation steps of factor analysis are basically the same. The difference lies in that the "rotation" dialog box in principal component analysis selects "none" while the "maximum variance method" in factor analysis.The eigenvalues and variance contribution values obtained are the same.The comprehensive factor score is calculated and ranked, and the result is the same as that of principal component analysis.

The comprehensive factor score is

$$F = 0.46936F1 + 0.16537F2 + 0.13997F3 + 0.13163F4 \quad (5)$$

According to the above chart, our team believes that among the 11 important indicators that can be measured digitally, the four key factors calculated are: tax, currency, employment and epidemic, which are the internal mechanisms affecting the US economy.

There are many ways to determine the weight. Based on the objective quantitative data, this paper adopts the objective weighting method to determine the weight.The specific calculation steps are as follows:

(1) Construct the judgment matrix X

Since the Election in the United States is held every four years, in order to make the data more timely, we need to select the eight years of data, that is, 8 evaluation objects and 11 indicators to choose the best evaluation object.Firstly, the judgment matrix $X = (X_{ij})_{8*11}$ is obtained through quantitative data statistics and other methods,and the $X_{ij}$ refers to the index value of the evaluation object i under the index j:

$$\begin{bmatrix} X_{11} & X_{12} & \dots & X_{1j} & \cdots & X_{1n} \\ X_{21} & X_{22} & \dots & X_{2j} & \dots & X_{2n} \\ \vdots & \vdots & \cdots & \vdots & \dots & \vdots \\ X_{i1} & X_{i2} & \dots & X_{ij} & \dots & X_{in} \\ \vdots & \vdots & \dots & \vdots & \dots & \vdots \\ X_{m1} & X_{m2} & \dots & X_{mj} & \dots & X_{mn} \end{bmatrix} \quad (6)$$

(2) Standardize the judgment matrix X

According to the "minimum-maximum standardization" formula, the judgment matrixX is standardized, and the judgment matrix $R = (r_{ij})_{m*n}$ is obtained.

$$\begin{bmatrix} r_{11} & r_{12} & \cdots & r_{1j} & \cdots & r_{1n} \\ r_{21} & r_{22} & \cdots & r_{2j} & \cdots & r_{2n} \\ \vdots & \vdots & \cdots & \vdots & \cdots & \vdots \\ r_{i1} & r_{i2} & \cdots & r_{ij} & \cdots & r_{in} \\ \vdots & \vdots & \cdots & \vdots & \cdots & \vdots \\ r_{m1} & r_{m2} & \cdots & r_{mj} & \cdots & r_{mn} \end{bmatrix} \quad (7)$$

(3) Calculate the entropy of information

fij refers to the proportion of the index value of i evaluation objects in the total value of the index of all evaluation objects in the index of j, as shown below,

$$f_{ij} = r_{ij}/\sum_{i=1}^{m} r_{ij} \ (m = 8, j = 1,2,...,11) \quad (8)$$

When fij = 0, ln (0) is meaningless, so they are amended to read:

$$f_{ij} = (1 + r_{ij})/\sum_{i=1}^{m}(1 + r_{ij}) \ (m = 8, j = 1,2,...,11) \quad (9)$$

$$H_j = -\frac{1}{\ln n}(\sum_{j=1}^{m} f_{ij} \ln f_{ij}) \quad (10)$$

(4) The index weight is determined according to the information entropy value
Calculate the entropy value of the JTH index and substitute it into Equation (15) to find the weight of the JTH index.

$$w_j = \frac{1-H_j}{m-\sum_{i=1}^{m} H_j} \ (\sum_{j=1}^{m} w_j = 1, m = 8) \quad (11)$$

Table 1. Weights of indicators for assessing the level of the US economy

| Indicators | The weight |
|---|---|
| Tax | 12.34% |
| Currency | 11.41% |
| Employment | 10.35% |
| Epidemic | 10.01% |
| Infrastructure | 8.31% |
| Education | 8.20% |
| Health care | 8.35% |
| Security | 8.20% |
| Science and technology | 8.14% |
| Financial regulation | 7.41% |
| Fiscal stimulus | 7.28% |

With so many differences between the two presidential candidates' policies[22][23][25], their priorities are also quite different. By consulting professional literature and consulting professional teachers, the policy tendencies of the two presidential candidates in the US fiscal policy, monetary policy, employment and livelihood, regulatory policy, epidemic situation and other aspects were obtained.

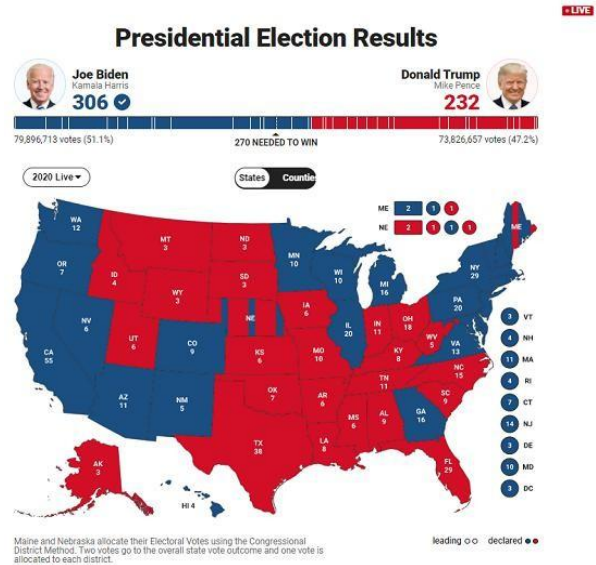

**Fig. 6.** Support for two candidates

As a result of the two presidential candidates in some aspects such as disease prevention and control measures for quantization can't use the data, we according to the global epidemic prevention and control of performance better national policy[28][30][33], and the two presidential candidates trump and biden fitting processing, epidemic prevention and control measures of and the results are normalized processing, do good to 1, do less to 0.

According to the above principles, the model results are divided into four grades: 0- 20 is very bad, 20-50 is poor, 50-80 is good, and 80-100 is very good (the grading thresholds of each grade include large or small), and the scoring and grading results of the two candidates are obtained.

**Table 2.** Weights of indicators for assessing the level of the US economy

| Candidate | Score | Grade | Rank |
|-----------|-------|-------|------|
| Trump | 0.622328 | 62.23 | Good |
| Joe Biden | 0.710164 | 71.02 | Good |

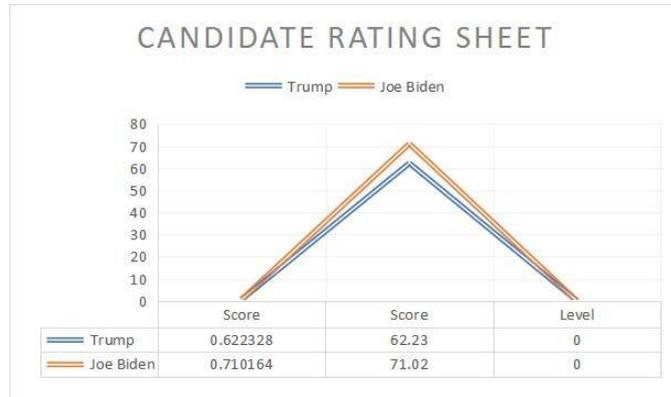

**Fig. 7.** Candiate Rating Sheet

As can be seen from the chart, both Trump and Biden have good ratings, but according to the specific value of the ratings, it can be seen that Biden's policy preference is more conducive to the development of the US economy.

On the basis of the establishment of the comprehensive evaluation model for the economic level of the United States, the possible impact of different candidates on China's economy is analyzed. According to the information searched, the 28 year data from 1992 to 2020 are collected. Some data are shown in the following table.

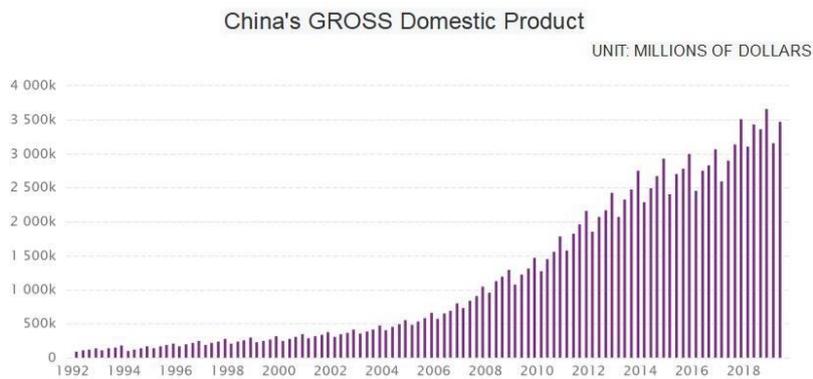

**Fig. 8.** 1992-2020 China's GROSS Domestic Product

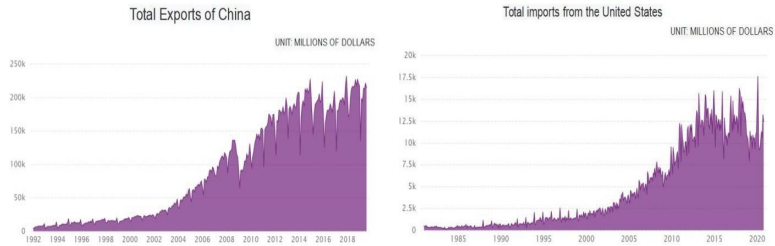

**Fig. 9.** Total Exports of China and Total imports from the United States

The index system constructed is optimized, and the "foreign policy" and other dimensional indicators are added to establish the index system as follows.

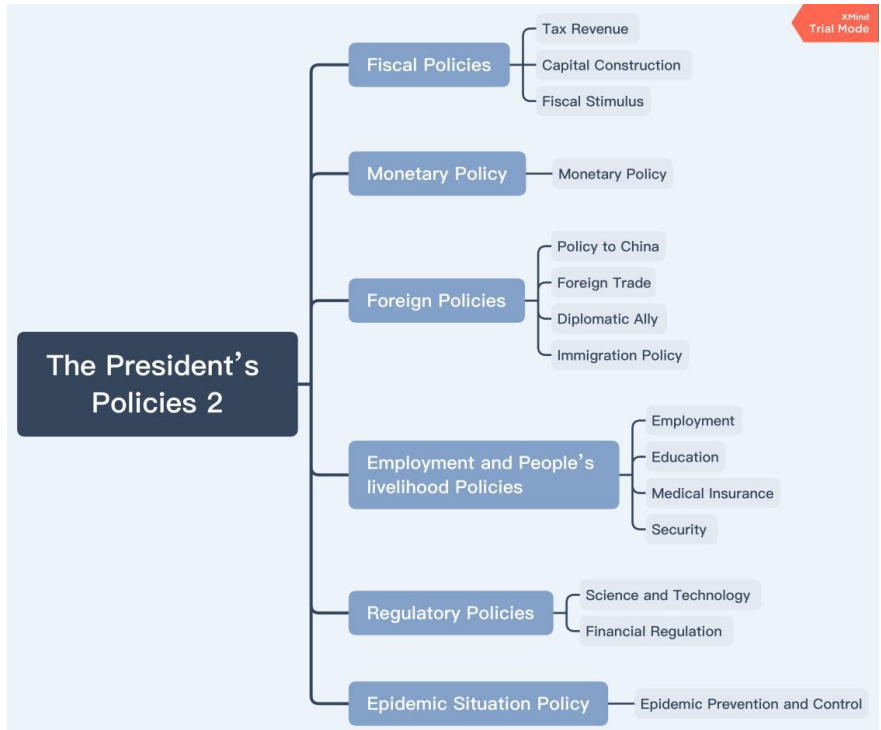

**Fig. 10.** The optimized index system is determined

Due to the difference of dimension and index value between the collected factors, it is necessary to use Excel for standardization. SPSS software was used to process the 15 factors by principal component analysis or factor analysis. The results of each factor were graded and analyzed by the model, and the specific quantitative index data of influence degree were obtained.

Due to the differences in the dimensions and indicator values of the collected factors, it is necessary to use Excel for standardization. SPSS software was used to analyze the 15 factors. The results of each factor are classified and analyzed by using the model, and the specific quantitative index data of influence degree are obtained. In other words, (max − min) standardizes the data and normalizes all index values into [0,1]. The formula is as follows:

$$R_{ij} = (X_{ij} - minX_j)/(maxX_j - minX_j) \qquad (12)$$

Among them, $minX_j$、$maxX_j$ are the minimum and maximum values of the j th index, and $R_{ij}$ is the standardized index value.

There are many methods to determine the weight, and the objective weighting method is also used to determine the weight.

(1) Construct judgment matrix X

In the study of the problem, because the election in the United States is held every four years, in order to make the data more timely, we need to select the best evaluation object among the eight evaluation objects and 15 indicators. Firstly, the judgment matrix is obtained by quantitative data statistics $X = (X_{ij})_{8*15}$ ($X_{ij}$ refers to the index value of the i evaluation object under the j index.

(2) Standardization of judgment matrix X

According to the formula of "minimum maximum standardization", the judgment matrix is normalized and the judgment matrix $R = (r_{ij})_{m*n}$ is obtained.

(3) Calculate information entropy

$f_{ij}$ refers to the proportion of the index value of the i evaluation object under the j index in the total value of the j index.

$$f_{ij} = r_{ij}/\sum_{i=1}^{m} r_{ij} \quad (m = 10, j = 1,2,\cdots,11\ m = 8, j = 1,2,3,\ldots,15) \qquad (13)$$

When $f_{ij} = 0$时，ln (0) is meaningless, so it was amended as follows:

$$f_{ij} = (1 + r_{ij})/\sum_{i=1}^{m}(1 + r_{ij}) \quad (m = 8, j = 1,2,3,\ldots,15) \qquad (14)$$

$$H_j = -\frac{1}{\ln n}(\sum_{j=1}^{m} f_{ij} \ln f_{ij}) \qquad (15)$$

(4) The index weight is determined according to the information entropy

The weight of j index can be calculated by introducing the entropy value of j index into formula (15).

$$w_j = \frac{1-H_j}{m-\sum_{j=1}^{m} H_j} \text{（其中，} \sum_{j=1}^{m} w_j = 1, m = 15 \text{）} \qquad (16)$$

The weight of China's economic evaluation index is calculated respectively, and the table is as follows:

Table 3. Weights of indicators for assessing the level of the US economy

| Index | Weight |
| --- | --- |
| Tax Revenue | 7.14% |
| Capital Construction | 7.12% |
| Fiscal Stimulus | 6.23% |
| Monetary Policy | 6.21% |
| Policy to China | 8.35% |
| Foreign Trade | 6.50% |
| Diplomatic Ally | 7.34% |
| Immigration Policy | 6.20% |
| Employment | 6.64% |
| Education | 6.41% |
| Medical Insurance | 6.33% |
| Security | 6.31% |
| Science and Technology | 6.58% |
| Financial Regulation | 6.28% |
| Epidemic Prevention and Control | 6.36% |

According to the above principles, the model results are divided into five grades (the critical value of each grade includes large but not small), and the scores and grading results of the two candidates are obtained.

Table 4. Weights of indicators for assessing the level of the US economy

| score | Grade |
| --- | --- |
| 0-20 | V |
| 20-40 | IV |
| 40-60 | III |
| 60-80 | II |
| 80-100 | I |

**Table 5.** Weights of indicators for assessing the level of the US economy

| Candidate | Score | Scoring | Grade |
|---|---|---|---|
| Trump | 0.6720 | 67.20 | II |
| Biden | 0.7692 | 76.92 | II |

It can be seen from the above chart that the two presidential candidates' inclination to govern the country has a great impact on China's economy. Among them, trump focuses on import and export trade with China and suppresses Chinese science and technology enterprises; Biden tends to restrict diplomacy and unite with many countries to boycott China.

# 5    Action plan for the US presidential election

Based on the model establishment and research analysis, we conclude that the challenges posed by the US election to China are as follows:

(1) Trade with the United States has declined sharply and industrial development has been constrained.The U.S. market has long been the focus of China's foreign trade.In 2017, China's exports to the US accounted for 19% of its total foreign trade and 4% of its GDP.China's traditional labor-intensive industries, such as toys, handicrafts, furniture and home textiles are heavily dependent on the US market, and the transformation of the export situation to the US has had a huge impact on these industries.As the trade war between China and the United States continues to heat up, once these traditional labor- intensive industries are put into the high-tariff basket, the supply of enterprises to the U.S. market will be greatly reduced.It is difficult for Chinese enterprises to find alternative markets in a short time, and their profits will fall off a cliff. China's foreign trade enterprises will face a more complex capital market environment, so the modernization and development of China's manufacturing industry is directly affected[31][32].

(2) RMB exchange rate falls and export competitiveness weakens.Since 2019, the RMB has been falling against the US dollar due to the appreciation of the US dollar and various trade wars.In 2020, the US added a $2 trillion stimulus bill due to the COVID-19 epidemic, bringing the foreign trade market into a deep freeze.The impact of exchange rate changes on foreign trade has both advantages and disadvantages. If RMB is devalued, theoretically enterprises can obtain higher profits for export, but at the same time, if raw materials needed by enterprises need to be imported, the corresponding import cost will also rise.The rise of RMB exchange rate leads to further rise of commodity prices, which inevitably affects the export volume of commodities and hinders the growth of exports. This will make our products less competitive compared with similar products of other countries, but also not conducive to our export.[35]

(3) The geographical direction of exports has changed, and new markets have brought new problems.The outbreak of the Trade war between China and the US, to some extent, limited the market share of the US export enterprises, reduced their revenues and reduced their scale, and many foreign trade enterprises had to shift their export objects to other markets in the world.Due to the regional differences, regional trade between the habit difference, come in different areas of consumer spending habits and preferences, such as Americans tend to flat color design firm of popular products, the Europeans prefer to bright and lively and rich design feeling, can reflect the quality of life of personalized products, this is for export enterprises production has brought certain difficulty, reducing the production efficiency, improving the production cost, new market brought new problems[28][31].

In response to the above challenges, the counter strategies are as follows:

(1) **Vigorously promote scientific and technological innovation and improve overall competitiveness**. The development of science and technology is the core of China's progress.During the trade war between China and the United States, Chinese enterprises were quite passive in the face of pressure from the United States. The main reason for this is that China's overall scientific and technological development level is still at a great disadvantage compared with that of some developed countries, and many key technologies and cutting-edge equipment are still in the hands of western developed countries.Core technology is mastered by foreign companies, the domestic enterprises can only be led by the nose.China is short of core technologies in the fields of medicine, precision manufacturing and military industry, and lacks a perfect patent protection system. As a result, China's sophisticated products have been monopolized by foreign companies for years.The West launched three industrial revolutions, and it took hundreds of years to accumulate today's scientific and technological achievements. Since the founding of the People's Republic of China, China has been rapidly catching up with western countries in the process of scientific and technological development.As far as today is concerned, Chinese enterprises should be more down-to-earth, increase investment in scientific research, actively transform to the high-tech industry, and assume social responsibility; The government should also encourage the development of high- tech enterprises and give them more support and help. Only in this way can Chinese enterprises substantially enhance their scientific and technological innovation capability, enhance their comprehensive competitiveness, have a sufficient say in foreign trade, and cope with various or current trade wars between China and the United States[26][27].

(2) **Continue to strengthen economic cooperation with developed countries**. To contain China's rise, the US is likely to join its traditional Allies, as evidenced by Mr Trump's customs sanctions on steel and aluminium.In response to this situation, first of all, While steadily advancing China's industrial modernization process, it is inevitable to have various conflicts with countries such as Europe and Japan, which are already in a high position.In this sense, China and its traditional Allies would also give up their natural interests in order to maintain the stability of the current international trading system. China should take this opportunity to actively maintain and develop friendly diplomatic relations with foreign developed economies. Many countries in the world have been cooperating with China to build domestic infrastructure, taking the lead in building 5G infrastructure with Huawei of China. The temperature between China and Japan has picked up. These facts clearly show that the United States cannot be the world's policeman and interfere with the development of world politics and economy. China needs to deepen cooperation with developed economies for mutual benefit, shoulder its responsibilities, and ensure a transparent and stable international free trade system[34][35].

(3) **Under the "One Belt And One Road" initiative, market diversification is realized**. The trade war between China and the United States has brought about changes in the overall foreign trade environment. In the trade conflict, China's foreign trade enterprises only to innovation as a vane, strive to adjust the direction of foreign trade, enhance their core competitiveness, cultivate and introduce excellent talents, pioneering and enterprising, can not go with the flow, to create their own oasis in the trade storm.China's foreign trade enterprises should develop new markets and actively adopt the strategy of export market diversification. The intrinsic motivation is to improve product quality, reduce product cost and upgrade the manufacturing industry. Within the framework of the belt and Road construction, China has carried out increasingly close economic and cultural communication with countries and regions along the Belt and Road, providing a trading platform for their development for mutual benefit and win-win results. Countries and regions along the Belt and Road naturally want to cooperate with us for a long time, thus generating a corresponding virtuous circle.Therefore, Chinese foreign trade enterprises have reduced their dependence on the American market and realized the diversification of the market. The national "One Belt And One Road" strategy has brought great opportunities for the development of China's foreign trade enterprises. Chinese foreign trade enterprises should seize the opportunity, actively cooperate with the national "One Belt And One Road" initiative, and seek better development under the framework of "One Belt And One Road"[27][30].

(4) **The government has accelerated financial support, enterprises have strengthened their own construction and actively expanded the domestic market**. In terms of labor and tax costs, China's foreign trade enterprises have certain disadvantages compared with neighboring developing countries, which is why some large international enterprises, such as Samsung, Nike and Adidas, have moved out of China to Vietnam and other countries.A variety of factors lead to the financing difficulties of China's foreign trade enterprises, especially small and micro enterprises, which cannot counter the economic losses brought by the trade war. This requires that our government must accelerate the improvement of fiscal and financial policies to reduce the difficulty of financing foreign trade enterprises and promote the development of foreign trade economy. Foreign trade enterprises should also firmly grasp the domestic market, export to domestic sales.Today, as the trade war between China and the United States continues to escalate, it is one of the best solutions for foreign trade enterprises to cope with the trade war between China and the United States by actively exploring the domestic market, tapping domestic consumer demand and keeping pace with domestic sales. At the same time, foreign trade enterprises should also actively create brands with international effect and make brands recognized by both domestic and foreign markets. In this way, they can better cultivate the domestic market, supply international and high-quality products to domestic consumers, and improve their ability to deal with foreign trade risks[18][23][31][35].